# Large field-like torque in amorphous $Ru_2Sn_3$ originated from the intrinsic spin Hall effect

Thomas J. Peterson[§,1], Mahendra DC[§,1], Yihong Fan[2], Junyang Chen[2], Delin Zhang[2], Hongshi Li[2], Przemyslaw Swatek,[2] Javier Garcia-Barriocanal[3], Jian-Ping Wang[1,2,*]

[1] School of Physics and Astronomy, University of Minnesota, Minneapolis, Minnesota 55455, USA

[2] Electrical and Computer Engineering Department, University of Minnesota, Minneapolis, Minnesota 55455, USA

[3] Characterization Facility, University of Minnesota, Minneapolis, Minnesota 55455, USA

* Email: jpwang@umn.edu; corresponding author

[§] Equal contribution authors

**Abstract**

We investigated temperature dependent current driven spin-orbit torques in magnetron sputtered $Ru_2Sn_3$ (4 and 10 nm) /$Co_{20}Fe_{60}B_{20}$ (5 nm) layered structures with in-plane magnetic anisotropy. The room temperature damping-like and field-like spin torque efficiencies of the amorphous $Ru_2Sn_3$ films were measured to be 0.14 ± 0.008 (0.07 ± 0.012) and -0.03 ± 0.006 (-0.20 ± 0.009), for the 4 (10 nm) films respectively, by utilizing the second harmonic Hall technique. The large field-like torque in the relatively thicker $Ru_2Sn_3$ (10 nm) thin film is unique compared to the traditional spin Hall materials interfaced with thick magnetic layers with in-plane magnetic anisotropy which typically have dominant damping-like and negligible field-like torques. Additionally, the observed room temperature field-like torque efficiency in $Ru_2Sn_3$ (10 nm)/CoFeB (5 nm) is up to three times larger than the damping-like torque (-0.20 ± 0.009 and 0.07 ± 0.012, respectively) and thirty times larger at 50 K (-0.29 ± 0.014 and 0.009 ± 0.017, respectively). The temperature dependence of the field-like torques show dominant contributions from the intrinsic spin Hall effect while the damping-like torques show dominate contributions from the extrinsic spin Hall effects, skew scattering and side jump. Through macro-spin calculations, we found that including field-like torques on the order or larger than the damping-like torque can reduce the switching critical current and decrease magnetization procession for a perpendicular ferromagnetic layer.

## Introduction

Devices designed with spin orbit torque (SOT) materials have been considerably studied as candidates for developing ultrafast-speed and ultralow-energy spin memory and logic applications, such as SOT magnetic random access memory (SOT-MRAM)[1–3]. The most commonly studied SOT generators are heavy metals (HMs), such as Ta[1,4,5]s , W[6–8], Pt[9–11], and topological insulators (TIs), such as $Bi_2Se_3$[12–18], $(Bi_xSb_y)_2Te_3$ [19,20], $Bi_2Te_3$[19]. HMs with low resistivities (10-300 $\mu\Omega cm$[1,4–11]) have charge-to-spin and spin-to-charge conversion efficiencies ($\zeta_S$) generated primarily by the bulk spin Hall effect, and have spin torque efficiency in the range of 0.01-0.5.[1,4–11] However, TI materials with larger resistivities (1000-100,000 $\mu\Omega cm$[12–20]) can have $\zeta_S$ much larger than 1. The efficient spin torque generation in TIs is due to spin polarized surface states where the electron's spin is perpendicularly locked with its momentum.[12,14,21–23] In addition to the spin-momentum locking, the bulk spin Hall effect[1,24] and the interfacial Rashba effect[9,25,26] can also generate spin torques in TI materials. The spin density generation in TIs has been shown to improve in nano-crystalline systems, sputtered $Bi_2Se_3$ has shown larger spin torque efficiencies (10-1000x) compared to single crystal molecular beam epitaxy grown $Bi_2Se_3$ due to quantum confinement from the grain size reduction in dimensionality.[15,17] Recent calculations on amorphous TI materials have shown that spin polarized edge states can exist and maintain topological protection even in fully amorphous systems.[27] Spin polarized surface states have been observed via angle resolved photoemission spectroscopy (ARPES) in amorphous $Bi_2Se_3$[28], and greater than 1 SOT efficiencies have been measured in amorphous Gd alloyed $Bi_2Se_3$.[29]

However, for SOT-MRAM applications the typical metallic free layers can have large current shunting due to the high resistivity of TI SOT channels, increasing the critical current required for magnetization switching. Novel lower resistivity topological materials are required to reduce the critical switching current density.[30] A possible new material is the $Ru_2Sn_3$ system. $Ru_2Sn_3$ is a low resistivity TI material, with a resistivity ranging from 800-2000 $\mu\Omega$cm (10x lower than sputtered $Bi_2Se_3$). The $Ru_2Sn_3$ band structure is a semiconductor at room temperature. Due to a crystalline phase change that occurs at 160 K, the band structure of the low temperature phase becomes a TI with highly anisotropic surface states.[31] The TI surface states in the low temperature phase have been observed through ARPES experiments.[31] The phase change is accompanied by a characteristic peak in resistivity centered at 160K.[32,33] The low temperature crystal phase has been experimentally shown to stabilize at room temperature via extreme applied pressures.[34] The $Ru_2Sn_3$ crystalline structure is also robust against annealing and stable up to 1100°C, making it a possible candidate for industrial application and CMOS integration.[35]

In this manuscript, we report large field-like torques with non-negligible damping-like torques in sputtered, amorphous $Ru_2Sn_3$ thin films. We confirm the amorphous structure of the films with high resolution tunneling electron microscopy and the 2:3 composition with Rutherford backscattering techniques. Through the second harmonic Hall measurement we can extract and characterize the effective damping-like (DL) and field-like (FL) SOTs originating in the $Ru_2Sn_3$ films. We find room temperature $\zeta_S^{DL}$ of 0.14 ± 0.008 and 0.07 ± 0.012 and a $\zeta_S^{FL}$ of -0.03 ± 0.006 and -0.20 ± 0.009 for 4 and 10 nm $Ru_2Sn_3$ films, respectively. By analyzing the resistivity dependence of the spin torque efficiencies, we extracted the contributions from the intrinsic and extrinsic spin Hall effects. The FL torques show dominant contributions from the intrinsic spin Hall effect with intrinsic spin conductivity while the DL torques show dominate contributions from the extrinsic spin hall effects, skew scattering and side jump. We performed macro-spin calculations of the Landau-Lifshitz-Gilbert (LLG) equation to simulate switching a magnetic layer including both the DL and FL torques generated from the $Ru_2Sn_3$ SOT channel. We find the inclusion of FL toques three times greater than the DL can reduce the critical current required for switching by almost 50% and reduce the magnetization precession compared to the DL torque only case.

## Materials Growth and Characterization

To investigate the SOTs in $Ru_2Sn_3$, thin films of $Ru_2Sn_3$(4 nm, 10 nm)/$Co_{20}Fe_{60}B_{20}$(5 nm) /MgO(2 nm)/Ta(2 nm) were deposited using magnetron sputtering on 300 nm thermally oxidized Si wafers, with a base pressure of 9 x $10^{-8}$ Torr and an Ar working pressure of 1.5 mTorr. The 4 and 10 nm $Ru_2Sn_3$ samples are labeled as RS4 and RS10, respectively. The $Ru_2Sn_3$ thin films were sputtered from a pure $Ru_2Sn_3$ alloy target but due to the non-empirical nature of magnetron sputtering, the composition may slightly drift from the optimal value. To confirm the ratio of Ru to Sn, the composition has been measured using Rutherford back scattering (RBS), with a He+ beam with maximum energy of 4.7 MeV and current of 40 μC, on a bare 17 nm $Ru_2Sn_3$ thin film, as shown in Supplemental Figure 8 (a). The final elemental composition is calculated from fitting the simulation results done in QUARK to the measured RBS data. The final composition of the sputtered $Ru_2Sn_3$ is calculated to be 40.2 and 59.8% (± 0.15%), respectively. Additionally, tunneling electron microscope (TEM) measurements were done on the RS4 and RS10 samples and show an amorphous phase of the $Ru_2Sn_3$. The bright-field TEM and High-Angle Annular Dark-Field (HAADF) images of RS10 shown in Figure 1 (a,b) suggests that no crystalline structure is observed in the Ru$_2$Sn$_3$ layer with small crystallites in the Co$_{20}$Fe$_{60}$B$_{20}$ and MgO layers. Further investigation including selected area diffraction did not reveal any long-range order in the Ru$_2$Sn$_3$ layer, indicating this layer is amorphous throughout, as shown in Figure 1 (c). Further measurements of XRD and Raman spectroscopy on 17 nm Ru$_2$Sn$_3$ thin films showed no indication of crystal structure, however, both measurements are limited by the film thickness.

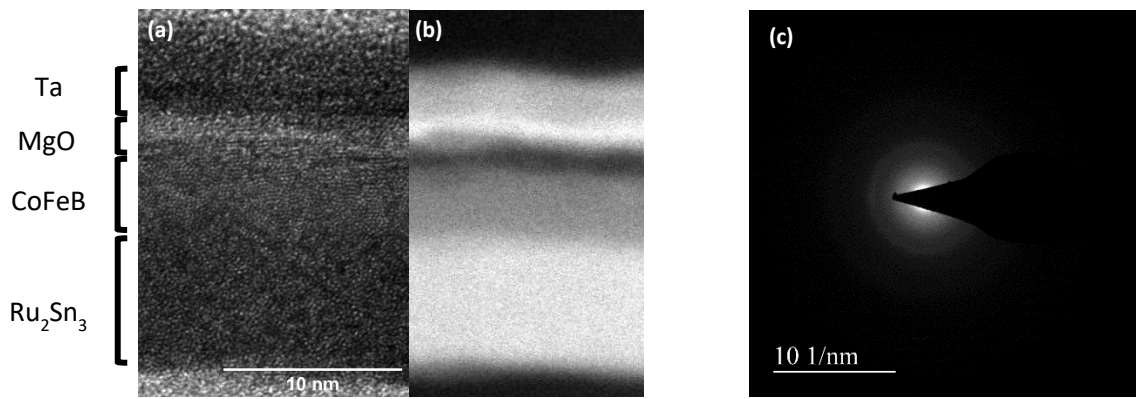

Figure 1. (a) Bright field transmission and (b) high-angle annular dark-field electron microscope image of the RS10 sample. (c) Selected area diffraction pattern of the RS10 sample. The diffuse ring seen suggests an amorphous film with no long-range order.

## Second Harmonic Hall Measurement for SOT Characterization

To calculate the charge to spin conversion efficiency of the DL and FL torque contributions independently we utilize the harmonic Hall measurement technique.[36–39] The RS4 and RS10 samples were patterned into Hall bars with a length of 85 μm and a width of 10 μm. An AC current with frequency 133 Hz and peak value of 4 mA is applied through the channel. Figure 2 (a) shows a schematic of the second harmonic measurement. The Hall bar is rotated in the xy plane from 0 to 360 degrees, while the first and second harmonic Hall voltages are measured via two lock in amplifiers. Figure 2 (b) shows the resulting first harmonic Hall voltage which provides the planar Hall resistance and can be fitted by:

$$V_{xy}^{\omega} = R_{PHE} \sin 2\varphi \, I, \tag{1}$$

where $\varphi$ is the in-plane angle. $R_{PHE}$ is the planar Hall resistance. Figure 2 (c) shows the second harmonic Hall voltages and is given by,

$$V_{xy}^{2\omega} = [V_{DL}\cos\varphi - V_{FL}(\cos 2\varphi \, \cos\varphi)]I \tag{2}$$

$$V_{DL} = \frac{1}{2} R_{AHE} \frac{-H_{DL}}{H_{ext} + H_k} + I \, \alpha \, \nabla T;$$

$$V_{FL} = R_{PHE} \frac{H_{FL} + H_{Oe}}{H_{ext}},$$

where $H_{ext}$ is the applied external field, $H_k$ is the perpendicular anisotropy field, and $I \, \alpha \, \nabla T$ is the field independent thermal contributions arising from the anomalous Nernst and Seebeck effects[37] (See Supplementary). $H_{Oe}$ is the Oersted field resulting from the current in the RS layer. $H_{Oe}$ is calculated to be roughly 0.12 and 0.43 Oe at 300K for the RS4 and RS10 samples, respectively.

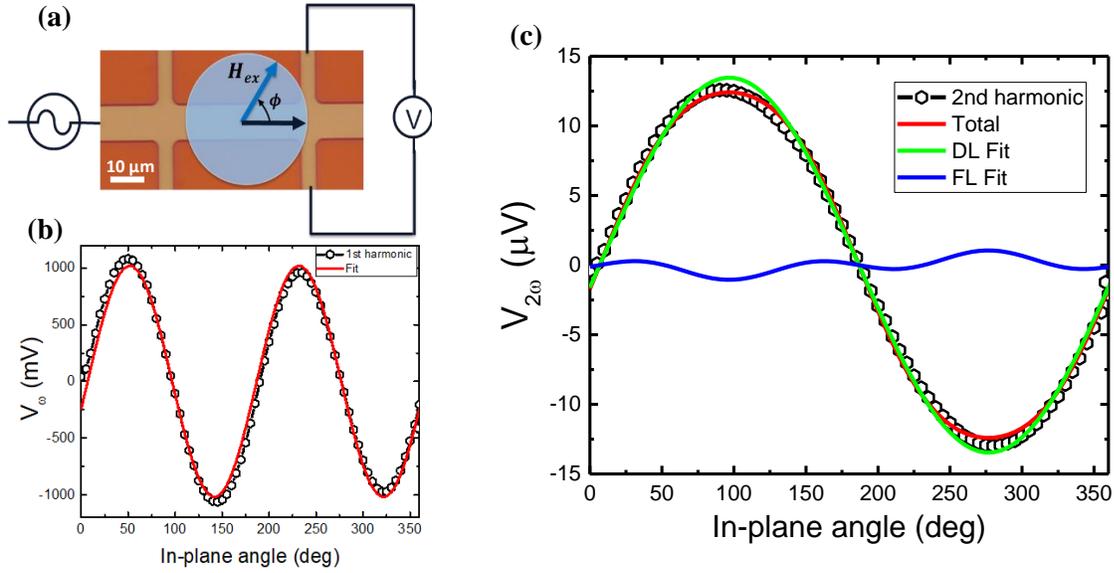

Figure 2. (a) Diagram of the harmonic Hall measurement, a Hall bar with a length of 85 μm and a width of 10 μm is rotated bar is rotated in the xy plane from 0 to 360 degrees, while the first and second harmonic Hall voltages are measured via two lock in amplifiers. (b) The first harmonic and (c) second Hall voltages for the RS10 sample rotated in a 1500 Oe external field at 300 K fitted to Equations 1 and 2, respectively.

There is a phase shift of roughly 90 degrees between the current direction and the field direction from sample mounting. $R_{AHE}$ is the anomalous Hall resistance and can be extracted by sweeping an out of plane field to 3 T and is measured down to 50K. The extracted 300K $R_{AHE}$ values are 9.5 and 7.9 Ω for the RS4 and 10 samples, respectively. The $V_{DL}$ and $V_{FL}$ voltages are extracted via fitting the second harmonic signals at various $H_{ext}$ from 0.15-3 T and at various sample temperatures from 50-300 K, shown in Figure 3 (a,b). $V_{DL}$ is fitted to a linear relation to separate the field independent thermal contributions from the DL-SOT. The thermal term is the dominant contribution to $V_{DL}$ in the samples ranging from 22 - 26 μV and 14 -18 μV for the RS4 and RS10 samples, respectively. The field dependence of $V_{FL}$ also has a non-zero intercept which does not fit the model in Equation 2, we have added an additional constant term to the linear fitting. The origin of this constant offset is still unclear, however, similar offsets have been seen in materials with large thermal contributions to the second harmonic signal such as Ta[37] and $W_xTe_{1-x}$.[40] The extracted value of $H_{FL}$ is about 5x larger than calculated value of $H_{Oe}$, suggesting the FL term is dominant over the Oersted contributions. $H_{DL}$ and $H_{FL}$ are the effective fields generated from the spin currents originating in the RS layer, and have the form:

$$H_{DL} = \frac{\hbar\, \zeta_{DL}\, J_{RuSn}}{2e\, M_S\, t_{CoFeB}} (\hat{\sigma} \times \hat{m}),$$

$$H_{FL} = \frac{\hbar\, \zeta_{FL}\, J_{RuSn}}{2e\, M_S\, t_{CoFeB}} [\hat{m} \times (\hat{\sigma} \times \hat{m})], \qquad (3)$$

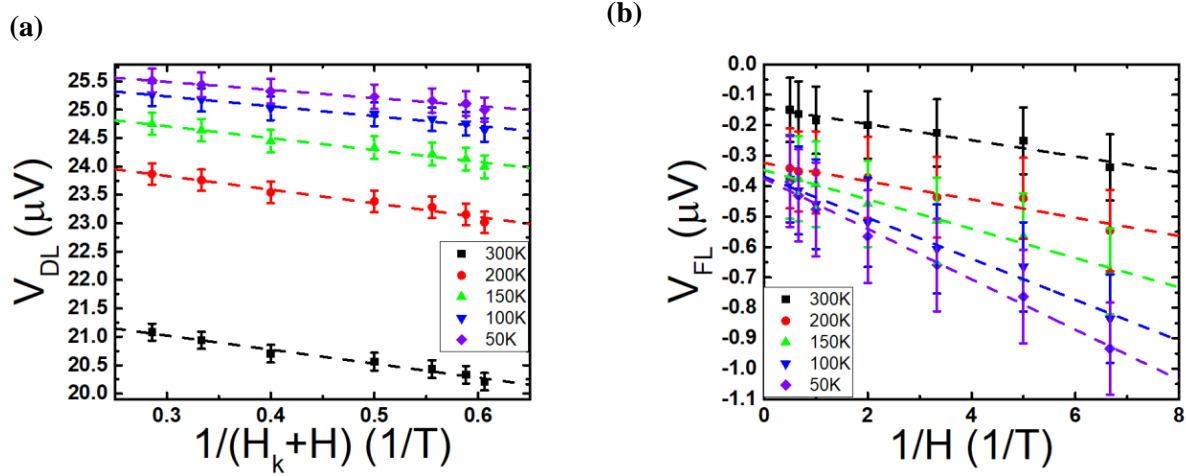

Figure 3. (a) The contribution from the DL and (b) FL torques on the second harmonic Hall voltage for the RS4 sample.

Where $\hat{\sigma}$ and $\hat{m}$ are the directions of the spin polarization and magnetization, respectively. $M_S$ is the saturation magnetization of the CoFeB layer and is measured via vibrating sample magnetometry down to 50K, shown in Supplemental Figure 7, the room temperature value of $M_s$ is 1100 emu/cc. $t_{CoFeB}$ is the thickness of the CoFeB layer and $J_{RuSn}$ is the estimated charge current flowing in the RS channel assuming parallel resistors. $\zeta_{DL}$ and $\zeta_{FL}$ are the effective charge to spin conversion efficiency for DL and FL torques, respectively. We depict the fitted values of $\zeta_{DL}$ and $\zeta_{FL}$, as a function of temperature in Figure 3 (a,b). Within a typical SOT material and FM bilayer the DL torque efficiency is from the vertical flow of spin current, whereas the FL torque efficiency are created from the accumulation of spin polarized electrons along the SOT/FM interface.

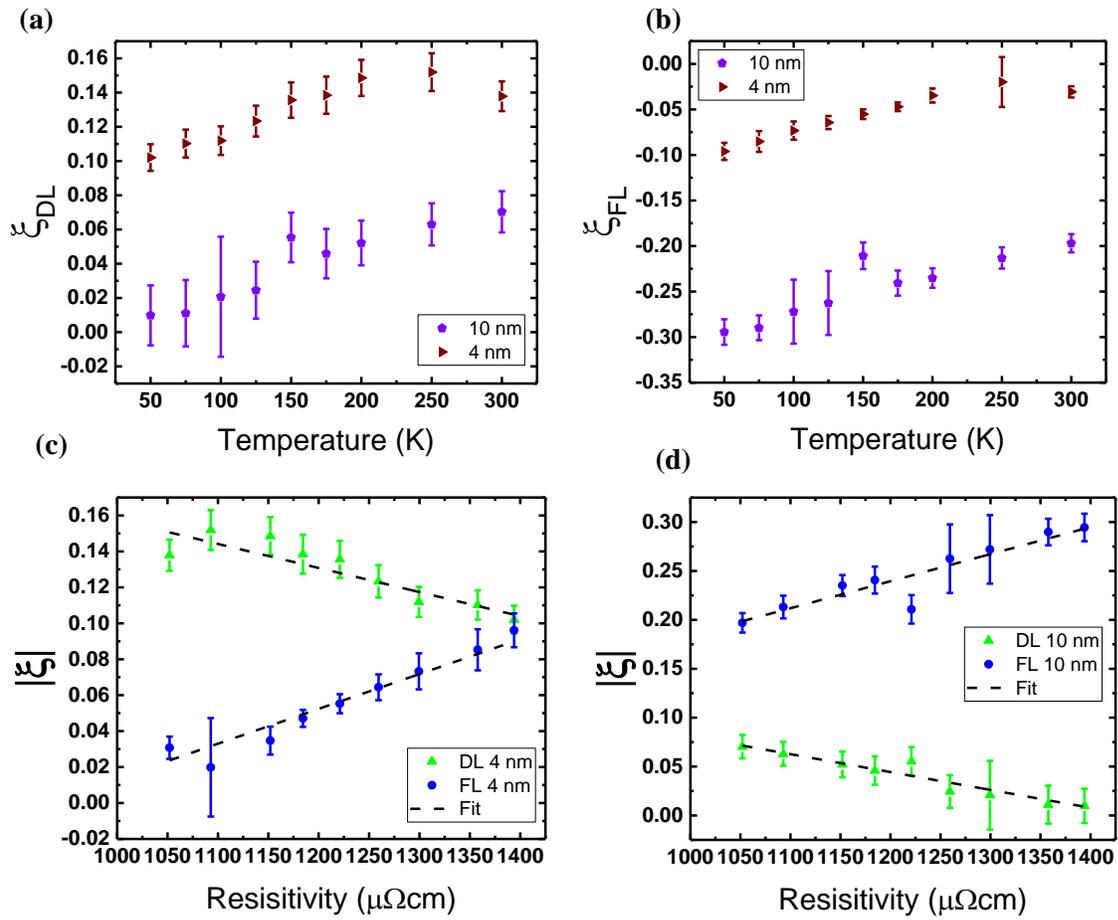

Figure 4. (a) Temperature dependence of the SOT efficiency from the DL torque and (b) FL torque. (c,d) Relation between magnitude $\zeta_{DL,FL}$ and the resistivity of the RS4 and RS10 samples, respectively. The dashed lines show the fit to Equation 4.

Although both types of torque in $Ru_2Sn_3$ exhibit nearly linear dependence on temperature, they follow opposite trends, i.e., the magnitude DL torque increases with increasing temperature, whereas the magnitude of the FL torque decreases with increasing temperature. The opposite behavior of DL and FL torques suggests the torques are originating from separate effects. The resistivity of the RS and CoFeB layers were estimated assuming a parallel circuit model, assuming the resistivity of the RS4 and 10 are the same, shown in Supplemental Figure 2 (b). The resistivity of the RS layer linearly increases with a decrease in temperature, without the characteristic peak in resistivity expected from the phase transformation expected at 160 K, indicating there was no crystalline transition into the TI phase. The temperature vs. resistivity observed suggests the films remained amorphous, however we still see large spin torques generated by the RS thin film, indicating the presence of large SOC. The extracted $\zeta_{DL}$ and $\zeta_{FL}$ also do not increase sharply below 160 K, which would be expected if the system transitioned into the TI phase due to the emergence of spin polarized topological surface states.

| | $\zeta$ | $\sigma_s^{Int}$ $(\Omega cm)^{-1}$ | $\sigma_s^{SJ}\rho_0^2 + \alpha_{ss}\rho_0$ $(\mu\Omega cm)$ |
|---|---|---|---|
| RS4 | DL | 13 ± 12 | -175 ± 19 |
| RS4 | FL | 117 ± 10 | -106 ± 15 |
| RS10 | DL | 73 ± 11 | -157 ± 16 |
| RS10 | FL | 240 ± 19 | -57 ± 27 |

Table I. Extracted intrinsic spin conductivity and extrinsic spin Hall resistivity components, side jump and skew scattering, from fitting equation 4 to the DL and FL spin torques for the RS4 and RS10 samples.

Figure 4 (c,d) shows the extracted spin torque efficiencies of the RS4 and RS10 samples as a function of the RS layer resistivity. The FL torque increases in magnitude from the low resistivity room temperature measurement to a higher value in the high resistivity low temperature measurement. However, the DL torques follow an opposite trend with resistivity suggesting the torques are arising from separate effects. The temperature dependent effective torque efficiencies can be analyzed by using following equation[41,42]

$$-\zeta = \sigma_s^{Int}\rho + \frac{\sigma_s^{SJ}\rho_0^2 + \alpha_{ss}\rho_0}{\rho}, \quad (4)$$

where $\rho$, $\rho_0$, $\sigma_s^{Int}$, $\sigma_s^{SJ}$, and $\alpha_{ss}$ are longitudinal resistivity, residual resistivity (resistivity at 0 K), intrinsic spin conductivity, spin conductivity due to the side jump, and skew scattering angle, respectively. Equation 4 assumes a transparent interface between the CFB and $Ru_2Sn_3$ and does not consider interface effects, such as spin memory loss and spin reflections. The residual resistivity of 1450 $\mu\Omega cm$ is extracted from the $Ru_2Sn_3$ resistivity vs temperature trend in Supplemental Figure 2 (b). As shown in Figure 3 (c,d), $\zeta_{FL}$ increases monotonically with $\rho$ indicating that the FL torque is originated from the intrinsic spin Hall effect. $\zeta_{FL}$ is larger in the RS10 compared to the RS4 sample which agrees with the intrinsic spin Hall effect mechanism. $\zeta_{DL}$ decreases with increase in $\rho$ against the belief of intrinsic spin Hall effect. By fitting Equation

4 to the data the estimated spin Hall parameters are extracted and shown in Table I. For both RS4 and RS10 the FL torque efficiencies show larger $\sigma_s^{Int}$ than their DL counterparts suggesting the FL torques are mainly originating from the intrinsic spin Hall effect. Similarly, the FL torques have smaller $\sigma_s^{SJ}\rho_0^2 + \alpha_{ss}\rho_0$ extrinsic parameters than the DL torques, suggesting the DL torques are originating from the extrinsic spin Hall effects.

The dominant FL torque term with non-negligible DL torque seen in the RS10 sample is an atypical result. Typically, HM systems with large SOC such as Pt, W, and Ta interfaced with thick magnetic layers with in-plane magnetic anisotropy are dominated by the SHE, generating a large DL and a negligible FL term.[37] In the case of Ta, non-negligible FL toques have been observed,[37,43,44] however, the DL torque typically remains the dominant torque. In the case of HM systems interfaced with thin magnetic layers with out of plane magnetic anisotropy FL can be comparable or larger than the DL torque.[4,43,45] Additionally, inserting a Hf spacer between W and a ferromagnet has been shown to increase the FL torque to be above the DL torque.[46] However, the FL torques in the RS films has very different temperature dependence than the Ta and W/Hf devices. In those samples the FL torque has a very strong temperature dependence, decreasing almost to zero and even changing sign of the FL torque at low temperatures while the damping-like torque is almost temperature independent.[43] The spin swapping effect has can also generate substantial FL torques and dominates in system where the disorder is high and spin orbit coupling is minimal.[47] The large DL torques observed in for the RS films suggests a strong spin orbit coupling, which does not fit with the spin swapping model. Large FL torques have been also observed in two-dimensional materials such as $MoS_2$ and $WSe_2$ due a strong Rashba-Edelstein effect; however, the DL torques in the monolayer systems are negligible since no bulk effects can contribute.[48] Bulk systems with strong interfacial Rashba torques, can cause an increase in FL torque efficiency due to the increase of bulk resistivity, increasing the current flowing through the interface increasing the FL torques. The decreased current through the bulk of the RS layer would decrease the current generated via the SHE reducing the DL torque efficiency. A similar resistivity dependence of the FL torque was observed in Ta systems.[43,44,49] However, in these interfacial Rashba systems the torques are generated at the interface and have minimal thickness dependence. The FL torques observed in the RS films have significant thickness dependence indicating a bulk material dominated effect.

To study the SOT switching of a magnetic layer considering large FL torques on the order or larger than the DL term, we performed macro-spin approximation calculations, see supplementary materials. The inclusion of large FL components in a simulated p-MTJ device reduces the magnetization precession and decreases the switching current by roughly 50%. For in-plane oriented magnetization we found the FL torque to have little to no impact on the switching dynamics.

## Conclusions

Current driven SOTs in magnetron sputtered $Ru_2Sn_3/Co_{20}Fe_{60}B_{20}$ layered structures with in-plane magnetization were investigated. We extracted the DL and FL spin torque efficiencies of the RS4 and RS10 samples utilizing the harmonic Hall technique. The calculated room temperature DL torque and FL torque for the RS4(RS10) system are 0.14 ± 0.008 (0.07 ± 0.012) and -0.03 ± 0.006 (-0.20 ± 0.009), respectively. The opposite trend of the DL and FL torques with temperature and resistivity suggest the torques are arising from separate effects. The FL torques show dominant contributions from the intrinsic SHE while the DL torques show dominate contributions from the extrinsic SHE. The trend of the torque efficiencies with temperature suggests the FL torque could also be arising from interfacial effects. We performed macro-spin calculations of switching a perpendicular and in-plane magnetization FM layer including contributions from both the DL and FL torques. The simulations suggest that including FL torques on the order or larger than the DL torque reduces $J_c$ and magnetization precession for perpendicular FM layers.

## Acknowledgements

This work was supported in part by ASCENT, one of six centers in JUMP, a Semiconductor Research Corporation (SRC) program sponsored by DARPA. This material is based upon work supported in part by the National Science Foundation under the Scalable Parallelism in the Extreme (SPX) Grant. The TEM was performed by Dr. Jason Myers and the RBS was performed by Dr. Greg Haugstad in the College of Science and Engineering (CSE) Characterization Facility at the University of Minnesota (UMN), supported in part by the NSF through the UMN MRSEC program. Portions of this work were conducted in the Minnesota Nano Center, which is supported by the National Science Foundation through the National Nano Coordinated Infrastructure Network (NNCI) under Award Number ECCS-2025124.